44

# Trends, Advancements and Challenges in Intelligent Optimization in Satellite Communication

Philippe Krajsic, Viola Suess, Zehong Cao, Ryszard Kowalczyk, Bogdan Franczyk

*Abstract*—Efficient satellite communications play an enormously important role in all of our daily lives. This includes the transmission of data for communication purposes, the operation of IoT applications or the provision of data for ground stations. More and more, AI-based methods are finding their way into these areas. This paper gives an overview of current research in the field of intelligent optimization of satellite communication. For this purpose, a text-mining based literature review was conducted and the identified papers were thematically clustered and analyzed. The identified clusters cover the main topics of routing, resource allocation and, load balancing. Through such a clustering of the literature in overarching topics, a structured analysis of the research papers was enabled, allowing the identification of latest technologies and approaches as well as research needs for intelligent optimization of satellite communication.

TABLE I
ABBREVIATIONS

| Abbr. | Term |
|---|---|
| ACO | Ant Colony Optimization |
| DDPG | Deep Deterministic Policy Gradient |
| DRL | Deep Reinforcement Learning |
| DRLR | Deep Reinforcement Learning-based Routing |
| ELM | Extreme Learning Machine |
| FANET | Flying Ad-Hoc Networks |
| GEO | Geostationary Orbit |
| IoT | Internet of Things |
| MANET | Mobile Ad-Hoc Networks |
| OT | Overarching Topics |
| QoS | Quality of Service |
| RCSA | Routing, Core, Spectrum Allocation |
| RL | Reinforcement Learning |
| RSA | Routing and Spectrum Allocation |
| RWA | Routing and Wavelength Allocation |
| SIoT | Satellite Internet of Things |
| TD3 | Twin Delayed Deep Deterministic Policy Gradient |

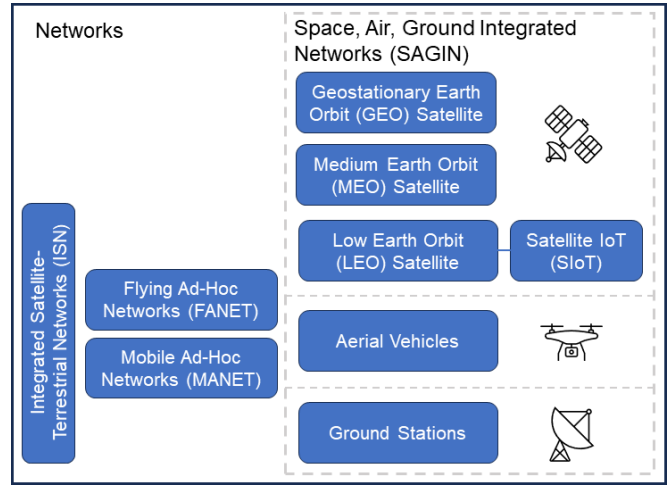

Fig. 1. Networks and Satellites.

## I. INTRODUCTION

Satellite communication plays a crucial role in modern communication systems, enabling global connectivity for a wide range of applications such as remote sensing [1], navigation [2], weather forecasting [3] or disaster management [4]. Over time, satellite communication systems have witnessed significant developments due to the increasing demand for reliable and high-bandwidth communication services, as well as rapid technological advancements. However, optimizing satellite communication systems remains an ongoing challenge despite these advancements. Optimization techniques in satellite communication aim to improve overall network performance by identifying and addressing potential issues and maximizing the efficiency of network resources. Optimization may involve the use of advanced signal processing algorithms, error correction techniques, or other methods to improve signal quality, increase data throughput, and reduce latency [5] [6].

New routing strategies are explored for Integrated Satellite-Terrestrial Networks (ISN). These are Space, Air, Ground Integrated Networks (SAGIN) including ground stations on earth, aerial vehicles like airplanes or drones as well as satellites in different orbit altitudes in space. These are Low Earth Orbit (LEO) satellites for short distances, Medium Earth Orbit (MEO) satellites, and Geostationary Orbit (GEO) satellites for long distances. Internet of Things (IoT) enabled sensors are often used to collect data in LEO satellite constellations, called Satellite IoT (SIoT) networks. In addition, ad-hoc network connections are established over Flying Ad-Hoc Networks (FANET), which may include Mobile Ad-Hoc Networks (MANET).

Previous surveys in this area often focus on the use of satellites in specific areas [7] or provide a more general overview of developments in the field of satellite communications [8]. The motivation for this literature review is to explore the current state of research in optimization in satellite communication systems. Specifically, we aim to identify the

Philippe Krajsic, Viola Suess and Bogdan Franczyk are with Leipzig University, Business Information System Institute, 04109 Leipzig, Germany (e-mail: krajsic@wifa.uni-leipzig.de, suess@wifa.uni-leipzig.de, franczyk@wifa.uni-leipzig.de). Zehong Cao and Ryszard Kowalczyk are with STEM, University of South Australia, Adelaide, SA 5095, Australia (e-mail: Jimmy.Cao@unisa.edu.au, Ryszard.Kowalczyk@unisa.edu.au). Corresponding author: Philippe Krajsic latest trends, advancements, challenges and potential research directions in this area of study. A comprehensive review of existing literature will provide insights into the most effective methods for optimizing satellite communication systems and guide future research directions. The importance of this review lies in the fact that satellite communication systems are critical to many industries and applications, and any improvements in their performance can have significant economic, social, and strategic benefits. Furthermore, as the demand for satellite communication services continues to grow, the optimization of the communication component across multiple satellites will become increasingly important. Therefore, it is crucial to have a clear understanding of the state-of-the-art research in this area and to identify the most promising venues for future research.

The leading research question that guides this review is: What are trends, advancements, and challenges in optimizing satellite communication systems? To answer this question, we will examine the existing literature on satellite communication systems. For this purpose, a structured, text-mining based, literature review was conducted. Search terms were defined with the help of which relevant research literature was searched in scientific databases. Subsequently, the identified literature was clustered using a text-mining approach and subdivided into thematic groups.

By synthesizing and analyzing the existing literature, we aim to provide insights into the most promising approaches for optimizing satellite communication systems and guide future research in this area.

The remainder of the paper is structured as follows: Section II presents the structured literature review process. Section III presents the findings of the systematic literature search. In section IV the synthesis of the identified literature takes place. Furthermore, the results of the literature review are discussed in section V. Possible limitations of the work are addressed in Section VI. The study concludes with a summary of all findings and an outlook on future work in section VII.

## II. STRUCTURED LITERATURE REVIEW PROCESS

In order to capture the current state of the art in the field of satellite communication optimization, methods have to be applied that allow for a structured review of this literature. One possible method is the structured literature review. According to [9], "A review of prior, relevant literature is an essential feature of any academic project. An effective review creates a firm foundation for advancing knowledge. It facilitates theory development, closes areas with a plethora of research exists, and uncovers areas where research is lacking." Such a review allows us to take a look at the current research and to filter out its theories and concepts. In the context of this review, a text mining based literature review was conducted based on the approach in [10]. This text mining based method enables researchers to systematically analyze large samples of research publications in an accessible way to enable a better understanding of this broad field of research. The individual steps of this procedure are shown in Figure 2.

The first step of the literature review is the definition of the review scope. This scope is derived from the guiding research questions formulated at the beginning and serves to narrow down the scope of the literature to be reviewed. Also contributing to the narrowing down is the definition of suitable search terms for the search in selected scientific databases. According to [11], the identification of keywords is an important step in the literature search process. To identify the keywords, the core questions "What?", "What for?" and "How?" represent three fundamental subject levels. Answering these questions leads to three core terms "Satellite" (What?), "Optimization" (What for?) and "Intelligent" (How?). On this basis, synonyms, generic terms and sub-terms can be defined, which are used to determine the search term. This search term is composed of individual components derived from the objective of the review. Keywords related to the considered satellite environment, the application of AI methods as well as terms in the context of optimization were included in these components. The individual components were linked with a Boolean 'AND' operator, and the elements within a component were also linked with an 'OR' operator:

TITLE-ABS-KEY ( "satellite" OR "satellite network*" OR "satellite system*" OR "hybrid satellite network*" OR "distributed satellite*") AND TITLE-ABS-KEY ( "load balancing" OR "routing" OR "resource optimi*ation" ) AND TITLE-ABS-KEY ( "artificial intelligence" OR "machine learning" OR "deep learning" OR "reinforcement learning")

The literature search was carried out based on the title, abstract and keywords of the publication in a period of five years, 2017-2023. The scientific databases used for the literature search were IEEE Xplore and Scopus, which provide extensive coverage of relevant literature in the research area under study. Additional inclusion criteria include publications exclusively in English in journals or at conferences. Thus, 176 publications and after applying the inclusion and exclusion criteria, as well as removing duplicates, 83 relevant publications were identified to be considered for the text mining based literature review.

For the text mining based literature review, the publications exported as CSV files were transferred to the data mining tool Orange [12]. Based on the provided functionalities in Orange, thematic clusters and superordinate topics can be extracted from the publications, which can then be used for a more in-depth analysis. The following steps were performed as part of the text mining analysis:

1) Corpus: Load a corpus of text documents. In this case the corpus is represented by the corresponding titles of the publications.
2) Preprocess Text: Preprocess text corpus with selected methods. Here, the pre-processor splits the text into tokens and normalizes the text of all titles.
3) Bag of Words: Generates bag of words from the input corpus. The bag of words function creates word counts for each instance. As a result rows and columns of entities and features are formed.
4) Distances: Computes distances of rows and columns in a dataset. Here, the Euclidian distances between the

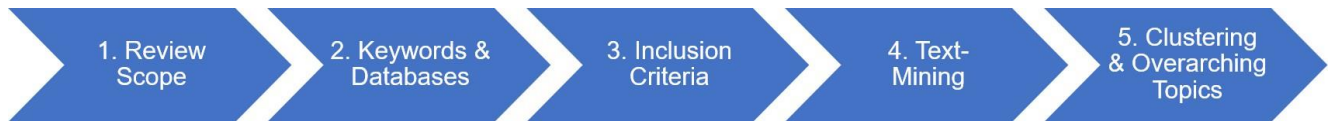

Fig. 2. Structured Literature Review Workflow.

rows of the dataset, resulting in a distance matrix, are computed.
5) Hierachical Clustering: Groups items using hierarchical clustering. This last step allocates all titles to different clusters.

## III. LITERATURE REVIEW RESULTS

Following the methodological approach outlined in the previous section, this paper provides an overview of research on optimzation in satellite communication published over the last five years.

The 83 relevant publications identified from the systematic literature review were published over a period from 2017-2023. The work has been published in a variety of publication venues, from conference proceedings to journal articles. It was found that the largest number of publications is held by the scientific conference IEEE Vehicular Technology Conference. Articles in these publication organs account for over half of the papers identified in this review. Using the text-mining workflow described in the previous section, clusters could be formed from the identified articles. The result of this clustering are the clusters shown in Table II. In total, the text mining workflow resulted in 20 thematic clusters to which the publications could be assigned. More homogeneous clusters were categorized as 'clear' (c), whereas clusters containing many titles with rather different focus were categorized as 'fuzzy' (f). The thematic clusters extracted by hierarchical clustering, were then grouped in three overarching topics (OT) manually based on the content of the studies. These OTs are Routing, Resource Allocation, and Load Balancing.

Most of the identified clusters have a clear thematic focus, although there are always overlaps in the assignment of OTs due to the thematic relationships. For example, for cluster C1 it can be seen that publications from the area of Routing and Resource Allocation optimization in satellite networks can be assigned to both OTs. Fewer clusters, on the other hand, are more fuzzy. One cluster, C20, which includes the most publications (22), contains a rather heterogeneous collection of topics, although this is a typical result when applying this text mining based approach.

These OTs differ in terms of their focus area, although there are also overlaps, which means that publications can be assigned to more than one OT in some cases (soft clustering). The following definitions, based on the identified literature, serve to better delimit the content of the defined OTs:

**Routing**: In the context of satellite communication, routing refers to the selection of the best path for data transmission between satellites and ground stations. Routing protocols for satellite networks aim to minimize signal delay and ensure reliable delivery of data between nodes [13] [14].

**Resource Allocation**: Resource allocation in satellite communication focuses on the efficient allocation of resources such as bandwidth, power, and frequency spectrum to ensure the best possible network performance. Effective resource allocation can help improve the capacity and efficiency of satellite communication networks and reduce the likelihood of congestion or network failures [15] [16].

**Load Balancing**: Load balancing techniques in satellite communication aim to distribute traffic across multiple satellites or ground stations to ensure that no single node is overloaded with traffic. Load balancing can help prevent congestion and ensure the smooth operation of satellite communication networks [17] [18].

In the following sections, the identified OTs are delineated from each other, with each OT being analyzed in detail further on.

## IV. SYNTHESIS

The following section includes the synthesis of the identified literature based on the mapping to the defined OTs Routing, Resource Allocation, and Load balancing.

### A. Routing

The first overarching research theme, Routing, consists out of sixteen thematic clusters: C1 Anti-Jamming, C2 Satellite Internet-of-Things, C3 Space-air-ground integrated networks, C4 Satellite-terrestrial networks, C5 Adaptive LEO satellite networks, C6 Ant Colony routing, C7 Reinforcement Learning (RL) based routing, C9 Machine learning based routing, C10 New routing methods, C11 Dynamic routing, C14 Software-defined satellite networks, C15 Inter-satellite networks, C16 Recovery routing, C17 Q-learning based communication, C18 RL based satellite communication, C19 Optimized routing and C20 Optimized satellite communication. This topic provides an overview of solutions for organizing satellite network topologies under constantly changing conditions for inter-satellite communications in conjunction with artificial intelligence methods to address the problems of time-limited ad hoc connections, low bandwidth, network interruptions, and unpredictable traffic spikes.

In the area of anti-jamming methods, covered by C1, a cross-layer anti-jamming method in satellite internet communication is proposed in [19]. Here, a Q-learning based anti-jamming algorithm considering routing delay, cost overhead

4TABLE II
THEMATIC CLUSTERS OF THE TEXT MINING BASED LITERATURE REVIEW.

| Cluster No | Cluster content | Clarity (c=clear, f=fuzzy) | OT |
|---|---|---|---|
| C1 | Anti-jamming | c | Routing |
| C2 | Satellite Internet-of-Things | c | Routing |
| C3 | Space-air-ground integrated networks | c | Routing, Resource Allocation, Load Balancing |
| C4 | Satellite-terrestrial networks | c | Routing, Load Balancing |
| C5 | Adaptive LEO satellite networks | c | Routing |
| C6 | Ant Colony routing | c | Routing, Load Balancing |
| C7 | Reinforcement Learning based routing | c | Routing, Load Balancing |
| C8 | Packet loss | c/f | Resource Allocation |
| C9 | Machine learning based routing | c | Routing |
| C10 | New routing methods | f | Routing |
| C11 | Dynamic routing | c/f | Routing, Load Balancing |
| C12 | Resource allocation | c | Resource Allocation |
| C13 | Federated learning | c | Resource Allocation |
| C14 | Software-defined satellite networks | c | Routing |
| C15 | Inter-satellite networks | c | Routing |
| C16 | Recovery routing | c | Routing |
| C17 | Q-learning based communication | c | Routing, Load Balancing |
| C18 | RL based satellite communication | f | Routing, Load Balancing |
| C19 | Optimized routing | f | Routing, Resource Allocation |
| C20 | Optimized satellite communication | f | Routing, Resource Allocation, Load Balancing |

and diversified jamming threats is presented. In addition a spatial anti-jamming scheme for internet of satellites, with the aim of minimizing anti-jamming routing cost via Stackelberg game and Deep Reinforcement Learning (DRL) is presented in [20]. One approach involves using a Deep Reinforcement Learning-based Routing (DRLR) Algorithm to optimize routing behavior in the Internet of Satellites [21]. In C2, SIoT is covered by the work in [22] [23] and [24]. Additionally, a Q-learning based reinforcement algorithm is used to develop an adaptive routing strategy for dynamically changing structures and states of agents in SIoT networks [22].

A gain-based limited greedy fast congestion control algorithm addresses transmission delays and interruptions in SIoT networks, reducing network interruptions in delay and interruption tolerant networks in [23]. Here, an algorithm for SIoT networks is presented using an improved ant lion optimization approach, respectively a method for optimizing resources in SIoT to address inter-cell interference and to maximize transmission capacity [24].

Work on SAGIN using Deep Learning is presented in [25]. A Deep Learning aided routing for space-air-ground integrated networks relying on real satellite, flight, and shipping data for supporting ubiquitous maritime communications using a Deep Reinforcement Learning aided multi-objective routing algorithm is presented in C3.

Routing strategies for ISN (C4) incorporate Convolutional Neural Networks, historical traffic data, fuzzy logic-based systems, and well-defined DRLR algorithms for better routing decisions and reduced delays [13] [26]. Finite-state Markov decision processes and Q-learning based routing algorithms optimize throughput, delay, and bit error rate by considering required resources and network constraints [27]. Space-time graph models, multipath routing, and minimum delay cost routing algorithms are developed to reduce routing delay and energy costs while providing efficient routing in LEO satellite networks [14].

The Approaches in C5 introduce an adaptive multipath routing algorithm for traffic planning in dynamic LEO satellite networks [28]. In [29] the redundancy according to the link quality is adaptively adjusted applying adaptive network coding algorithm based on hop-by-hop feedback.

Ant Colony Optimization (ACO) mechanisms are employed in several works in C6 to optimize satellite network routing by finding the optimal path and improving Quality of Service (QoS). These include Q-learning based algorithms that consider dynamic topology, feedback from individual nodes, and multipath routing algorithms to achieve optimal path selection [30] [31] [32].

Deep Reinforcement Learning algorithms (C7) with Q-learning trained models enable intelligent next hop selection in [33] [34] [35] [36]. Other optimization techniques include decentralized load balancing [17], Extreme Learning Machines (ELMs) for traffic predictions [37].

Machine learning-based algorithms for routing optimization (C9) in mega-satellite constellations is a topic in [38]. In [39] an extended extreme learning machine algorithm, which can predict the communication attenuation caused by rainy weather to satellite communication links to avoid large path loss caused by bad weather conditions is presented.

A paper presents new routing methods [C10] for LEO satellite networks aiming at high quality, efficiency, and capacity by using QoS-based artificial intelligence algorithm routing algorithm and a multi-layer satellite network routing algorithm [40].

Dynamic routing (C11, C14) adaptation problems are addressed through battery-saving routing algorithms, fully distributed dynamic routing, multi-commodity flow routing, and intelligent routing patterns using Software Defined Networking in [41] [42] [43] [44] [45] [46].

In particular, the work [47] [48] focuses on the optimization of inter-satellite networks (C15) in LEO networks for dynamic next-hop planning with routing-based algorithms. Based on fault detection with Reinforcement Learning algorithm, a new satellite network routing technology for fault recovery (C16) is proposed in [49].

In work [50] a delay-aware routing algorithm based on Q-learning (C17) is proposed. Here, swarms of aerial vehicles communicating using 6G technology are used to send data to the next hop in LEO networks.

Satellite communication based on multi-agent Reinforcement Learning, ACO metaheuristics, and Q-learning algorithms are utilized to optimize energy consumption, inter-satellite communication, and routing in satellite delay tolerant networks in [51] [52] [53]. Machine learning-based classification algorithms predict potential neighbor nodes in [54] for optimized routing (C18, C19).

In C20 different topics concerning optimized satellite communication are covered. There are resilient routing strategies, adaptive snapshot routing strategies, and distributed routing with spiking neural networks provide solutions for various routing challenges in [55] [56] [57] [58] [59]. Other research areas include secure and robust communication in satellite networks [60], adaptive routing in FANETs [57], delay-free packet loss tolerant networks [58], optimal timing and location planning of tankers [54], and intelligent routing decisions based on Q-learning algorithms [61]. Machine learning algorithms are used to design load balancing communication [62], find robust data transmission links [63], and develop optimal routing strategies [64]. Furthermore, ant pheromone-based Markov's decision process, Reinforcement Learning, flooding and greedy algorithms, and balancing link distance and hop count are applied to multipath traffic scheduling, integrated routing mechanisms, routing balancing, and wavelength assignment algorithms [65] [66] [67] [68]. In [58], [69], [70] and [23] reinforcement respectively Deep Reinforcement Learning approaches are applied to solve several challenges, like optimal cost-latency satellite data downloading, latency minimization and energy optimization, key node identification and congestion control. Furthermore, ACO as an extended path-finding routing and wavelength assignment algorithm is used in [71] to improve global search capability.

**Potential research directions**

From a scientific point of view, OT routing, especially routing optimization, in satellite networks offers further possible research directions. This literature review focuses on AI-based routing optimization from machine learning to Reinforcement Learning algorithms, with emerging categories of solutions to the routing problems of QoS improvement, satellite topology modofication, and hardware power consumption optimization. This includes the adjustment of anti-jamming methods utilizing learning and reasoning [19], the combination of centralized scheduling and distributed game to further improve the performance of congestion control in the satellite internet-of-things scenarios [23] the improvement of optimization schemes for the routing design of space-air-ground integrated network architectures [39] or the convergence of IoT, internet and satellite networks [24]. In [28] [30] [31] [32] [54] [57] [68] [24] [71] ACO mechanisms are used to optimize the satellite routing problem. For example, there is still room to explore agent-based Deep Reinforcement Learning, which is only addressed in [51], [52], and [42].

## B. Resource Allocation

In OT Resource Allocation, overlapping topics can also be found. Resource allocation can be found in the clusters of Space, Air, Ground Integrated-Networks (C3), Packet loss (C8), Resource allocation (C12), Federated Learning (C13), Optimized routing (C19) and Optimized satellite communication (C20).

In [66] a routing scheme based on Reinforcement Learning aiming at the limited energy and bandwidth of different nodes and congestion in the SAGIN is proposed. Here, a RL based algorithm is implemented that has a low overhead and can adjust routing in real time (C3). C8 concentrates on Packet loss. Using AI-optimized traffic channel adjustments, [72] simulates a packet loss scenario in SAGIN. A Deep Reinforcement Learning based algorithm based on the state-of-the-art Twin Delayed Deep Deterministic Policy Gradient (TD3) to jointly allocate the subchannel and power for dynamic resource allocation is presented in [73] in C12. Therefore, the problem is modeled as a Markov Decision problem and solved using a Deep Reinforcement Learning approach based on TD3. In [15] a short-message satellite resource allocation algorithm based on Deep Reinforcement Learning is proposed. Therefore, a Deep Reinforcement Learning algorithm based on the Deep Deterministic Policy Gradient (DDPG) framework is used to improve the resource utilization of short-message satellite communication system and ensure the service quality. An overview on routing and resource allocation based on machine learning algorithms in optical networks is provided in [71]. Here, machine learning enabled RWA (routing and wavelength allocation) algorithms, RSA (routing and spectrum allocation) algorithms, and RCSA (routing, core, spectrum allocation) algorithms are elaborated, analyzed and compared. In [5] (C13) uses mobile edge computation and federated learning for satellite modulation detection.The problem is solved using a Q-learning based algorithm. In the area of Optimized routing (C19) an optimal scheduling for the location geosynchronous satellites refueling problem is presented in [54]. In this work the scheduling problem arising from refueling multiple geosynchronous satellites with multiple servicing vehicles is addressed applying an ACO algorithm. The work in [74] and [59] is concerned with meta-learning for Optimized satellite communication (C20). In [74] a meta-critic learning based efficient resource scheduling is presented. In [59] the problem of data pre-storing and routing in dynamic, resource-constrained cube satellite networks using distributed and distribution-robust meta Reinforcement Learning is studied. An energy efficient approach to maximal throughput resource scheduling in satellite networks is shown in [16]. Furthermore, an overview of beam hopping algorithms (meta heuristic algorithms, Deep Reinforcement Learning based algorithms, heuristic algorithms) in large scale LEO satellite constellation is given in [75].

**Potential research directions**

From a scientific point of view, the topic of Resource Allocation opens up some interesting further potential research directions. These include, for example, cooperative routing between multiple layer networks [66], a unified scheduling method of time–space-frequency multidimensional resources [76], a continuous learning and behavior regularization, to further improve sample efficiency and model adaptability [74], the investigation of link discontinuity and limited transceiver and power resources as well as the implementation of a high dynamic topology of satellite networks [16].

*C. Load Balancing*

The OT Load Balancing covers popular topics, all covered by other OTs as well, like Space, Air, Ground Integrated-Networks (C3), Satellite-terrestrial networks (C4), topics in the intersection with other OTs, like Ant Colony routing (C6), Reinforcement Learning based routing (C7), Dynamic routing (C11), Q-learning based communication (C17), Reinforcement Learning based satellite communication (C18) and Optimized satellite communication (C20).

In C3 a Q-learning based traffic scheduling scheme for load balancing in software defined network based space-air-ground-integrated-networks is presented in [77]. The proposed approach enhances the transmission capability of SAGIN with a software defined network architecture using a Deep Reinforcement Learning model to make global optimal traffic scheduling decisions. In [78] a method for outsourcing partial computations from the satellite to the ground station is developed for balanced hardware utilization with DNN optimization models. In [18] a fuzzy-logic based load balancing scheme for a ISN (C4) is presented that proposes a fuzzy evaluation metric to pre-evaluate the users impact on overload. Here, an adaptive neuro fuzzy system based on Reinforcement Learning is introduced to implement the load balancing scheme. In [79] a probabilistic ACO routing (C6) algorithm with window reduction for LEO satellite networks is proposed to achieve load balancing. The suggested approach limits the movement of the Ant Colony to a specific range and comprehensively considers the path distance, transmission direction, and link load to find a path with low delay and overhead. An intelligent decentralized load balancing routing algorithm using Deep Reinforcement Learning (C7) for the LEO satellite networks is proposed in [17]. A Deep Q-network algorithm takes into consideration the queuing delay, storage space, communication bandwidth and propagation delay to learn a routing strategy by utilizing the status of only one-hop neighboring satellite nodes. Paper [75] investigates beam-hopping algorithms for Dynamic routing (C11) in LEO satellite networks to improve system throughput with DRL. (C17) A multi-agent Reinforcement Learning based load balance architecture using Q-Learning for optimization communication is presented in [80]. Here, the load balance in satellites communications is formulated as a partially observable Markov decision process and uses a collaborative Q-learning algorithm to search for the best load balancing strategy. Reinforcement Learning as well as Deep Reinforcement Learning are frequently used optimization techniques applied to load balancing problems. For RL based satellite communication (C18) in [81] a satellite network traffic scheduling algorithm based on multi-agent Reinforcement Learning is used. Machine learning technologies continue to make their way into the area of load balancing. A Machine learning based frequency resource demand prediction for optimized mobile satellite communication (C20) in a satellite network is proposed in [82]. Here, an architecture and employment of a link-layer user-uplink simulator and a methodology that enables prediction of the spectrum demand of various connection service mixtures applying random forest algorithm is presented. Reviewing literature in the field of load balancing, it can be seen that, as in the previous OTs, algorithms from the field of machine learning, in particular from the fields of Reinforcement Learning and Deep Reinforcement Learning, are gaining in importance.

**Potential research directions**

As has been shown, the improvement of load balancing strategies in satellite networks is an essential aspect in the optimization of satellite communications. From the literature reviewed, further potential research directions can be identified. This includes the further improvement of high-rate communications in remote areas with sparsely distributed terrestrial cells [18].Further research directions include the investigation of energy consumption in load balancing algorithms for satellite networks [79], an effective utilization of inter satellite links [80] and the development of efficient communication techniques for multi-layer networks [82]. Due to the widespread use of Reinforcement Learning techniques, the further utilization of multi-agent Deep Reinforcement Learning approaches for load balancing challenges also needs to be studied.

V. DISCUSSION

This literature review was motivated by the question which trends, advancements and challenges exist in the field of intelligent optimization of small sized satellite communication systems. To address this question, a systematic, text-mining-based literature review was conducted for the period 2017-2023. Using a clustering approach, relevant articles were thematically summarized and sorted into overarching topic groups of Routing, Resource Allocation, and Load Balancing. These topic clusters differ in their thematic focus, but by their nature also have overlaps, which can lead to assignments of papers to more than one thematic cluster. Within the framework of the study, the following insights were gained:

Several trends can be observed in the research on intelligent optimization of satellite communication. First, there is a clear trend towards the application of Reinforcement Learning and Deep Reinforcement Learning algorithms. RL-based routing schemes are proposed to address challenges such as limited energy and bandwidth, congestion, and resource allocation in satellite networks. DRL algorithms based on the TD3 and DDPG frameworks are employed for subchannel and power allocation in dynamic resource management. Additionally, meta-learning approaches are explored for efficient resource scheduling. The adoption of these Machine learning techniques reflects the growing importance of artificial intelligence for optimization in satellite routing, resource allocation, and load balancing.

TABLE III
ML ALGORITHMS USED IN SATELLITE TECHNOLOGY.

| Learning approach | Learning techniques |
|---|---|
| Supervised Learning | Bayesian Decision Theory, Convolutional Neural Networks, Deep Neural Network, Extreme Learning Machine, Federated Learning, K-Nearest Neighbors, Naive Bayes, Spiking Neural Network, Support Vector Machine |
| Unsupervised Learning | Long-Short Term Memory Network |
| Reinforcement Learning | Deep Deterministic Policy Gradient, Distributed Distribution-Robust Meta RL, Markov Decision Process, Multi Agent RL, Q-Learning |

Another trend is the transfer of existing algorithms such as ACO algorithms in satellite networks. It is applied to solve routing and scheduling problems, such as the optimal scheduling for refueling GEO satellites and routing optimization for LEO satellite networks. These optimization algorithms provide effective solutions for resource allocation and routing in satellite communication systems.

Furthermore, Machine learning algorithms are employed in optical networks for routing and resource allocation. ML-enabled algorithms, including RWA, RSA, and RCSA, are investigated, analyzed, and compared. This trend highlights the importance of machine learning techniques in improving the efficiency and performance of optical satellite communication networks.

The advancements in intelligent optimization of satellite communication are notable. One significant advancement is the application of RL and DRL algorithms for real-time routing and resource allocation. These algorithms provide adaptive and efficient solutions to dynamic and resource-constrained satellite networks. The use of DRL algorithms based on TD3 and DDPG frameworks enables better resource utilization and ensures service quality in short-message satellite communication systems.

It was found that multi-factorial optimization must be considered in the three OTs to meet the challenges of satellite communication. Table III breaks down the machine learning based methods and algorithms used in the literature review. It is clear that supervised learning and reinforcement learning are a common area of research in satellite communications optimization.

Moreover, meta-learning approaches have shown promise in optimizing satellite communication. Meta-critic learning is employed to improve resource scheduling efficiency, while distributed and distribution-robust meta Reinforcement Learning techniques are used for data pre-storing and routing in dynamic, resource-constrained cube satellite networks. These advancements contribute to the development of intelligent and adaptable satellite communication systems.

Despite the significant advancements, several challenges remain in the field of intelligent optimization of satellite communication. One of the challenges is the need for cooperative routing between multiple layer networks. Integrating satellite communication with other communication layers, such as terrestrial networks, poses challenges in terms of resource allocation and coordination.

Another challenge lies in the effective utilization of limited resources, including hardware, transceiver power, and bandwidth. The effective utilization of limited resources, is a crucial challenge in satellite communication systems. Maximizing resource utilization while maintaining service quality and minimizing congestion requires intelligent optimization algorithms.

In addition, the dynamic and unpredictable nature of satellite communication networks presents challenges for routing and resource allocation. Satellite networks are subject to various environmental factors and changing network conditions, which require adaptive and real-time decision-making algorithms. Developing efficient algorithms that can adjust routing and resource allocation in real-time is essential to ensure optimal performance.

Furthermore, energy efficiency is a critical challenge in satellite communication. Satellite networks often operate with limited energy resources, and optimizing energy consumption while maintaining communication quality is a complex task. Addressing this challenge requires the development of energy-efficient algorithms and optimization strategies.

Interference and congestion management is another significant challenge in satellite communication systems. As the demand for satellite services increases, the risk of interference and congestion rises, leading to degraded performance and reduced service quality. Designing intelligent algorithms and strategies to mitigate interference and effectively manage network congestion is crucial for optimizing satellite communication systems.

Moreover, the scalability and complexity of large-scale satellite constellations pose challenges for optimization. As satellite constellations grow in size and complexity, the optimization algorithms need to scale accordingly to handle the increasing number of satellites and the dynamic network topology. Developing efficient optimization techniques that can handle the complexities of large-scale satellite constellations is an ongoing challenge.

Lastly, the integration of machine learning algorithms into satellite communication systems introduces the need for robust and interpretable models. Machine learning techniques, such as RL and DRL, offer promising solutions for optimization problems. However, ensuring the reliability, interpretability, and robustness of these models remains a challenge. Addressing this challenge requires the development of robust and explainable machine learning models that can be effectively applied to satellite communication systems.

While significant advancements have been made in the field of intelligent optimization of satellite communication, several challenges remain. Addressing these challenges, such as cooperative routing, resource utilization, dynamic decision-making, energy efficiency, interference management, scalability, and model robustness, will drive further research and development in the field of intelligent optimization of satellite communication systems.

## VI. LIMITATIONS

Studies such as these are subject to certain limitations. These limitations are determined by the defined methodology for a structured literature review, the defined publication period, and the scientific databases used. In addition, inclusion and exclusion criteria are incorporated into the selection of literature, minimizing the size of the literature base. In certain cases, these restrictions can lead to a non-inclusion of relevant literature.

## VII. CONCLUSION

The presented study aimed to investigate current literature in the field of satellite communication optimization. For this purpose, a structured literature review was first conducted to identify relevant papers from the research area under investigation. Using a text-mining based search, thematic clusters were formed, which were then grouped into overarching topics. These OTs, Routing, Resource Allocation, and Load Balancing, were then analyzed for their contribution to the optimization of satellite communication. Besides the analysis of technologies and methods, potential research directions for future innovations could be identified. The results obtained provide an overview of the current state of the literature in the field of optimization of satellite communication using modern technologies, in particular from the field of artificial intelligence.

Future research work aims to further exploit the results obtained. This includes in particular the development of intelligent methods, such as agent-based Deep Reinforcement Learning approaches, for load balancing strategies for LEO satellites.

## ACKNOWLEDGMENTS

This work was supported by the German Academic Exchange Service (DAAD - 57653780) for funding Programmes for Project-Related Personal Exchange Australia 2023-2025.